\documentclass[a4paper,11pt]{article}
\usepackage[reqno,fleqn]{amsmath}
\usepackage{graphicx}
\usepackage[T1]{fontenc}
\author {Hossein.Mehraban\footnote{hmehraban@semnan.ac.ir}, {Behnam.Mohammadi\footnote{$b_{-}$mohammadi@sun.semnan.ac.ir}}\\Physics Department, Semnan University\\P.O.Box 35195-363, Semnan, Iran}
\title{Measurement of the Quasi-Two-Body B Decays}
\begin{document}
\maketitle
\begin{abstract}
We study the contributions of the $B\rightarrow
\psi(3770)K[\psi(3770)\rightarrow D\bar{D}]$, $B\rightarrow
K^*(1410)\pi[K^*(1410)\rightarrow K\pi]$ and $B\rightarrow
X(3872)K[X(3872)\rightarrow J/\psi\gamma, \psi(2S)\gamma,
D\bar{D}\pi, J/\psi\omega, J/\psi\pi\pi$ and $D\bar{D}^*\pi]$
quasi-two-body decays. There are no existing previous measurement
of the three-body branching fractions for three final states of
the $X(3872)\rightarrow J/\psi\gamma$, $\psi(2S)\gamma$ and
$D\bar{D}\pi$ but several quasi-two-body modes that can decay to
this final state have been seen.
\end{abstract}

\section{Introduction}
The three-body meson decays are generally dominated by
intermediate vector and scalar resonances, this means that, there
are a resonance state and a pseudoscalar meson which they proceed
by quasi-two-body decays \cite{R.H,J.P,H.Y,B.Au}. In fact,
analysis of three-body B decays using the Dalitz plot technique
leads us to the many quasi-two B decays \cite{R.H}. The study of
the $B\rightarrow \psi(3770)K$, $B\rightarrow K^*(1410)\pi$ and
$B\rightarrow X(3872)K$ via quasi-two-body decays were considered.
These include $B^+\rightarrow \psi(3770)K^+$ and $B^0\rightarrow
\psi(3770)K^0$, observed in the $\psi(3770)\rightarrow
D^0\bar{D}^0$ channel and also seen in $\psi(3770)\rightarrow
D^+D^-$; $B^0\rightarrow K^*(1410)^+\pi^-$, seen in
$K^*(1410)^+\rightarrow K^0\pi^+$; $B^+\rightarrow X(3872)K^+$ and
$B^0\rightarrow X(3872)K^0$, seen in $X(3872)\rightarrow
J/\psi\pi^+\pi^-$, $J/\psi\omega$ and $\bar{D}^{*0}D^0K^0$ channel
\cite{J.B}. The decays $B^+\rightarrow X(3872)K^+$ and
$B^0\rightarrow X(3872)K^0$ have also been observed with
$X(3872)\rightarrow J/\psi\gamma$, $\psi(2S)\gamma$ and
$D^0\bar{D}^0\pi^0$ \cite{J.B}. In this article, we do not perform
a Dalitz plot analysis, but instead use information on
intermediate modes including narrow resonances by studying the
two-body invariant mass distributions, because there are no
existing previous measurement branching fractions for some of the
three-body decays such as $X(3872)\rightarrow J/\psi\gamma$,
$\psi(2S)\gamma$ and $D^0\bar{D}^0\pi^0$, but quasi-two-body modes
that can decay to these final states have been seen. Hence we
present the results of a search for the three-body decay including
short-lived intermediate two-body modes that can decay to these
final states.\\
In general factorization approach, to obtain the amplitudes of the
two-body decays, the Feynman quark diagrams should be plotted,
quasi-two-body decays of the heavy mesons can be also expressed in
terms of some quark-graph amplitudes. For example we take
$B^0\rightarrow K^*(1410)^+\pi^-$ as an illustration. Under the
factorization approach, its decay amplitude consists of three
distinct factorizable terms: (i) the current-current process
through the tree $b\rightarrow u$ transition, (ii) the transition
process induced by $b\rightarrow s$ penguins and (iii) the
annihilation process. Note that weak-annihilation contributions
are too small so we ignore them in our calculations.

\section{Quasi-Two-Body Decay Amplitudes}
\begin{figure}[t]
\begin{center} \includegraphics[scale=.8]{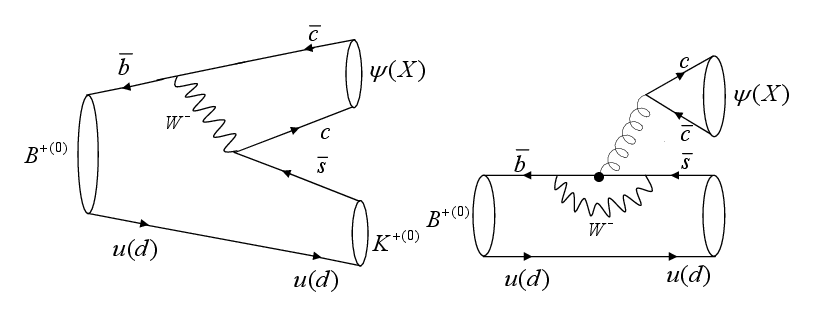}
\caption{\label{fig:b1}Quark diagrams illustration the processes
of the $B^{+(0)}\rightarrow \psi K^{+(0)}$ and $XK^{+(0)}$
decays.}\end{center}
\end{figure}

\begin{figure}[t]
\begin{center} \includegraphics[scale=.8]{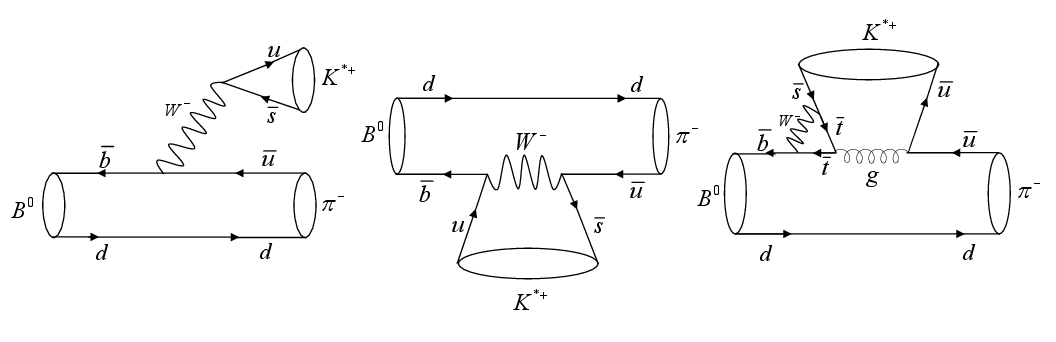}
\caption{\label{fig:b2}Quark diagrams illustration the process of
the $B^0\rightarrow K^{*+}\pi^-$ decay.}\end{center}
\end{figure}

It is known that in the narrow width approximation, in the models
we use to obtain the amplitudes of the decays, the 3-body decay
rate obeys the factorization relation \cite{J.P}
\begin{eqnarray}\label{eq1}
Br(B\rightarrow RM\rightarrow M_1M_2M)=Br(B\rightarrow RM)\times
Br(R\rightarrow M_1M_2),
\end{eqnarray}
with R being a vector meson resonance and $M$, $M_1$ and $M_2$ are
pseudoscalar and vector final state mesons. The intermediate
vector meson contributions to three-body decays are identified
through the vector current, their effects are described in terms
of the Breit-Wigner formalism. The Breit-Wigner resonant term
associated to  quasi two body $R+M$ state which seems to play an
important role as indicated by experiments. We have to calculate
the branching ratios of the $Br(B\rightarrow RM)$ by using the
Feynman quark diagrams and use the experimental information for
the $Br(R\rightarrow M_1M_2)$ decays as follows \cite{J.B}:
\begin{eqnarray}\label{eq2}
Br(\psi(3770)\rightarrow D^0\bar{D}^0)&=&(52\pm5)\%\nonumber\\
Br(\psi(3770)\rightarrow D^+D^-)&=&(41\pm4)\%\nonumber\\
Br(K^*(1410)^+\rightarrow K^0\pi^+)&=&(6.6\pm1.3)\%\nonumber\\
Br(X(3872)\rightarrow J/\psi\pi^+\pi^-)&>&2.6\%\nonumber\\
Br(X(3872)\rightarrow J/\psi\omega)&>&1.9\%\nonumber\\
Br(X(3872)\rightarrow J/\psi\gamma)&>&9\times10^{-3}\nonumber\\
Br(X(3872)\rightarrow D^0\bar{D}^0\pi^0)&>&3.2\times10^{-3}\nonumber\\
Br(X(3872)\rightarrow \bar{D}^{*0}D^0)&>&5\times10^{-3}\nonumber\\
Br(X(3872)\rightarrow \psi(2S)\gamma)&>&3.0\%.
\end{eqnarray}
We calculate the branching ratios of the intermediate states
two-body decays. Feynman diagrams related to these decays are
shown in Figs. \ref{fig:b1} and \ref{fig:b2}.
A detailed discussion of the QCD factorization (QCDF)  approach
can be found in \cite{A.A1,A.A2,M.B1,M.B2}. Factorization is a
property of the heavy-quark limit, in which we assume that the b
quark mass is parametrically large. The QCDF formalism allows us
to compute systematically the matrix elements of the effective
weak Hamiltonian in the heavy-quark limit for certain two-body
final states $\psi K^{+(0)}$, $X K^{+(0)}$ and $K^{*+}\pi^-$. In
this section, we obtain the amplitude of $B^{+(0)}\rightarrow \psi
K^{+(0)}$, $B^{+(0)}\rightarrow X K^{+(0)}$ and $B^0\rightarrow
K^{*+}\pi^-$ decays by using the QCDF method. We adopt leading
order Wilson coefficients at the scale $m_b$ for QCDF approach.
According to the QCDF, the amplitudes of the $B^{+(0)}\rightarrow
\psi K^{+(0)}$, $B^{+(0)}\rightarrow X K^{+(0)}$ and
$B^0\rightarrow K^{*+}\pi^-$ decays are given by
\begin{eqnarray}\label{eq3}
A(B^{+(0)}\rightarrow \psi(3770)
K^{+(0)})&=&\sqrt{2}G_{F}m_{\psi}f_\psi
(\epsilon_\psi.p_B)F_{1}^{BK}(m_\psi^2)\nonumber\\
&&\times(a_2V_{cb}V_{cs}^{*}+a_3\lambda_p),
\end{eqnarray}
\begin{eqnarray}
A(B^{+(0)}\rightarrow X(3872) K^{+(0)})&=&\sqrt{2}G_{F}m_{X}f_X
(\epsilon_X.p_B)F_{1}^{BK}(m_X^2)\nonumber\\
&&\times(a_2V_{cb}V_{cs}^{*}+a_3\lambda_p),
\end{eqnarray}
\begin{eqnarray}\label{eq5}
A(B^0\rightarrow K^*(1410)^+\pi^-)&=&\sqrt{2}G_{F}m_{K^*}f_{K^*}
(\epsilon_{K^*}.p_B)F_{1}^{B\pi}(m_{K^*}^2)\nonumber\\
&&\times[(a_1+a_2)V_{ub}V_{us}^{*}+a_4\lambda_p],
\end{eqnarray}
where
\begin{eqnarray}\label{eq6}
\epsilon_V.p_B&=&\frac{m_B}{m_V}|\vec{p}_V|, (V=\psi, X, K^*)\nonumber\\
\lambda_{p}&=&\sum_{p=u,c}V_{pb}V^{*}_{ps},\nonumber\\
F_{1}^{BP}(q^2)&=&\frac{F(0)}{1-a_F(q^2/m_B^2)+b_F(q^2/m_B^2)^2},
(P=K, \pi).
\end{eqnarray}
where $|\vec{p}_V|=\sqrt{(p^0_V)^2-(m_V)^2}$ is the absolute value
of the 3-momentum of the vector meson in the B rest frame.
\section{Numerical results}
To proceed with the numerical calculations, we need to specify the
input parameters. For the CKM matrix elements, we use
$V_{u_b}=0.00351^{+0.00015}_{-0.00014}$,
$V_{u_s}=0.22534\pm0.00065$, $V_{c_b}=0.0412^{+0.001}_{-0.0005}$
and $V_{c_s}=0.97344\pm0.00016$ \cite{J.B}. For $B\rightarrow K$
and $B\rightarrow \pi$ form factors, a good parametrization for
the $q^2$ dependence can be given in terms of three parameters
(see Eq. (\ref{eq6})). We fix for $B\rightarrow K$ transition
$F(0)=0.374$, $a_F=1.42$, $b_F=0.434$ \cite{P.B} and for
$B\rightarrow \pi$ transition $F(0)=0.25$, $a_F=1.73$, $b_F=0.95$
\cite{H.Y2}, namely, $F^{BK}(m_\psi^2)=0.962$,
$F^{BK}(m_X^2)=1.283$ and $F^{B\pi}(m_{K^*(1410)}^2)=0.284$. The
meson masses and decay constants needed in our calculations are
(in units of Mev) $m_B=5279.25\pm0.17$, $m_\psi=3772.15\pm0.33$,
$m_X=3871.68\pm0.17$, $m_{K^*(1410)}=1414\pm15$ \cite{J.B};
$f_\psi=384\pm14$ \cite{F.M}; $f_X=446\pm16$ \cite{R.D}. The
Wilson coefficients $c_i$ have been calculated in different
schemes. In this paper we will use consistently the naive
dimensional regularization(NDR) scheme. The values of $c_i$ at the
scale $\mu=m_b$ at the leading order (LO) and next to leading
order (NLO) are shown in table \ref{tab2} \cite{M.B2,G.B}.
\begin{table}[t]
\caption{\label{tab2}Wilson coefficients $c_i$ in the NDR scheme
at the leading order and next to leading order.}
\begin{tabular}{|c|c|c|c|c|}
  \hline
   & $c_1$ & $c_2$ & $c_3$ & $c_4$ \\
  \hline
  NLO & 1.081 & -0.190 & 0.014 & -0.036 \\
  LO & 1.117 & -0.268 & 0.012 & -0.027 \\
  \hline
\end{tabular}
\end{table}
Numerical values of effective coefficients $a_i$ for
$\bar{b}\rightarrow \bar{s}$ transition at $N_c=3$ are shown in
table \ref{tab3}.
\begin{table}[t]
\caption{\label{tab3}Numerical values of effective coefficients
$a_i$ for $\bar{b}\rightarrow \bar{s}$ transition at the leading
and next to leading order and $N_c=3$. }
\begin{tabular}{|c|c|c|c|c|}
  \hline
   & $a_1$ & $a_2$ & $a_3$ & $a_4$ \\
  \hline
  NLO & 1.02 & 0.17 & 0.003 & -0.02 \\
  LO & 1.03 & 0.11 & 0.002 & -0.03 \\
  \hline
\end{tabular}
\end{table}
According to table \ref{tab3}, it is not much difference between
effective coefficients $a_i$ at the LO and NLO,
therefore we use them at the LO: $a_1=1.03$, $a_2=0.11,$, $a_3=0.002$ and $a_4=-0.03$ \cite{M.B2,G.B}.\\
Now we are able to calculate the branching ratios of the
$B^{+(0)}\rightarrow \psi(3770) K^{+(0)}$, $B^{+(0)}\rightarrow
X(3872) K^{+(0)}$ and $B^0\rightarrow K^*(1410)^+\pi^-$ decays by
using the Eqs. (\ref{eq3})-(\ref{eq5}) as follows 
\begin{eqnarray}\label{eq7}
Br(B^{+(0)}\rightarrow \psi(3770)
K^{+(0)})&=&(2.95^{+1.10}_{-0.86})\times10^{-4}\nonumber\\
Br(B^{+(0)}\rightarrow
X(3872) K^{+(0)})&=&(3.12^{+1.00}_{-0.85})\times10^{-4}\nonumber\\
Br(B^0\rightarrow
K^*(1410)^+\pi^-)&=&(2.57^{+0.58}_{-0.52})\times10^{-5}
\end{eqnarray}
The experimental result for $Br(B\rightarrow \psi(3770) K)$ which
turns out to be $(4.9\pm1.3)\times10^{-4}$ \cite{J.B} in very good
agreement with our prediction. As we know the branching ratio of
$B\rightarrow X(3872) K$ decay has already been estimated: a)\;in
\cite{C.M} they have used QCD sum rules to calculate the branching
ratio and predicted $(0.38\pm0.06)\times10^{-6}$, b)\;in
\cite{E.B} they have analyzed the decay $B\rightarrow XK$ and the
decays of B into $D^0\bar{D}^{0*}K$ and $D^{0*}\bar{D}^0K$ near
the threshold for the charm mesons by separating the decay
amplitudes into short-distance factors and long-distance factors,
they have predicted for branching ratio $2.9\times 10^{-5}$ while
the experimental result of this decay is $Br(B\rightarrow
\psi(3770) K)<3.2\times 10^{-4}$ \cite{B.A}. By using the Eqs.
(\ref{eq1}), (\ref{eq2}) and (\ref{eq7}) the branching ratios of
quasi-two-body decays are summarized in table \ref{tab1}. Note
that the experimental BRs for X(3872) decay modes (see Eq.
(\ref{eq2})) only have lower bounds, so the theory predictions
involving these modes in table \ref{tab1} only have lower bounds.

\begin{table}[t]
\caption{\label{tab1}Branching ratios of quasi-two-body decays
obtained from the studies of three-body decays based on the
factorization approach.}
\begin{tabular}{|c c|c|c|c|}
  \hline
   & Decay mode & Theory & Exp. \cite{J.B} & Units \\
  \hline\hline
   \rule{0pt}{4ex}$\psi(3770)K^+$; & $\psi\rightarrow D^0\bar{D}^0$ & $1.56\pm0.32$ & $1.6\pm0.4$ & $\times10^{-4}$ \\
    & $\psi\rightarrow D^+D^-$ & $12.28\pm2.61$ & $9.4\pm3.5$ & $\times10^{-5}$ \\
    \hline\hline
   \rule{0pt}{4ex}$\psi(3770)K^0$; & $\psi\rightarrow D^0\bar{D}^0$ & $1.56\pm0.32$ & $<1.23$ & $\times10^{-4}$ \\
    & $\psi\rightarrow D^+D^-$ & $1.23\pm0.26$ & $<1.88$ & $\times10^{-4}$ \\
    \hline\hline
    \rule{0pt}{4ex}& $X\rightarrow J/\psi\pi^+\pi^-$ & $>9.17$ & $8.6\pm0.8$ & $\times10^{-6}$ \\
    & $X\rightarrow J/\psi\gamma$ & $>3.20$ & $2.1\pm0.4$ & $\times10^{-6}$ \\
   $X(3872)K^+$; & $X\rightarrow \psi(2S)\gamma$ & $>10.59$ & $4\pm4$ & $\times10^{-6}$ \\
    & $X\rightarrow D^0\bar{D}^0\pi^0$ & $>0.016$ & $1.0\pm0.4$ & $\times10^{-4}$ \\
    & $X\rightarrow \bar{D}^{0*}D^0$ & $>0.20$ & $8.5\pm2.6$ & $\times10^{-5}$ \\
    \hline\hline
    \rule{0pt}{4ex}& $X\rightarrow J/\psi\pi^+\pi^-$ & $>9.17$ & $4.3\pm1.3$ & $\times10^{-6}$ \\
    & $X\rightarrow J/\psi\gamma$ & $>3.20$ & $<2.41$ & $\times10^{-6}$ \\
   $X(3872)K^0$; & $X\rightarrow \psi(2S)\gamma$ & $>1.06$ & $<6.61$ & $\times10^{-6}$ \\
    & $X\rightarrow D^0\bar{D}^0\pi^0$ & $>0.016$ & $1.7\pm0.8$ & $\times10^{-4}$ \\
    & $X\rightarrow \bar{D}^{0*}D^0$ & $>0.20$ & $1.2\pm0.4$ & $\times10^{-4}$ \\
    & $X\rightarrow J/\psi\omega$ & $>6.72$ & $6.0\pm3.2$ & $\times10^{-6}$ \\
  \hline\hline
   \rule{0pt}{4ex}$K^*(1410)^+\pi^-$; & $K^*\rightarrow K^0\pi^+$ & $2.53\pm0.21$ & $<3.8$ & $\times10^{-6}$ \\
  \hline
\end{tabular}
\end{table}

\section{Conclusion}
In this research  we have calculated the branching ratios of the
$B\rightarrow \psi(3770)\\K[\psi(3770)\rightarrow D\bar{D}]$,
$B\rightarrow K^*(1410)\pi[K^*(1410)\rightarrow K\pi]$ and
$B\rightarrow X(3872)K$ $[X(3872)\rightarrow J/\psi\gamma,
\psi(2S)\gamma, D\bar{D}\pi, J/\psi\omega, J/\psi\pi\pi$ and
$D\bar{D}^*\pi]$ decays in the framework of the quasi-two-body
method. We have also measured the branching ratios of two-body
decays including the short-lived intermediate mesons by using the
QCDF method. Our calculation results are shown in table
\ref{tab1}. There are no existing previous measurement branching
fractions for some of the three-body decays such as $B \rightarrow
J/\psi\gamma K$, $\psi(2S)\gamma K$ and $D\bar{D}\pi K$, but
quasi-two-body modes that can decay to these final states have
been seen.

\end{document}